
\magnification=\magstep1
\hsize=14.6 truecm
\vsize=21.5 truecm
\baselineskip=12 pt
\def\pt{\partial_\tau}
\def\ps{\partial_\sigma}
\def\Pt{{P\over2}\tau}
\def\boxit#1#2{\vbox{\hrule\hbox{\vrule
\vbox spread#1{\vfil\hbox spread#1{\hfil#2\hfil}\vfil}%
\vrule}\hrule}}
\def\dal{\boxit{6pt}{$\!$}\, }
\null \vfill
\centerline {{\bf The Nappi-Witten string in the light-cone gauge}
\footnote{$^\dagger$}{to appear in the E.P. Wigner memorial volume
of Acta Physica Hungarica}}
\vskip 1.5truecm
\centerline{P. Forg\'acs}
\centerline{Research Institute for Particle and Nuclear Physics}
\centerline{H-1525 Budapest 114, P.O.B. 49, Hungary}
\vskip .5truecm
\centerline{P.A. Horv\'athy}
\centerline{D\'epartement de Math\'ematiques}
\centerline{Universit\'e de Tours}
\centerline{ Parc de Grandmont, F-37200 TOURS, France}
\vskip .5truecm
\centerline{Z. Horv\'ath and L. Palla}
\centerline{Institute for Theoretical Physics}
\centerline{Roland E\"otv\"os University}
\centerline{H-1088 Budapest, Puskin u. 5-7, Hungary}
\vskip 4.5truecm

\centerline{{\bf Abstract}}
\bigskip

Some of the motivations for as well as the main points of
the quantization of the
Nappi Witten string in the light cone gauge are reviewed.
\vfill\eject

Recently the problem of strings propagating in various classical
field backgrounds has received an increased interest [1-5]. The Nappi Witten
string is a particular example of this type: it describes the propagation
of a bosonic string in a particular gravitational wave and axion field
background. These classical fields were obtained in [3] from an
ungauged WZW model based on $E_2^c$, the central extension of the two
dimensional Euclidean group. In [3] it was also shown, that these background
fields solve the one loop $\beta$-function equations.

In [4] the highest weight
representations of the Kac-Moody (KM) algebra corresponding to this WZW
model were studied and it was pointed out that there are three types of
Kac-Moody representations with unitary base. These three possibilities
correspond to the three different unitary representations of the
zero grade algebra (ZGA) (which is nothing but the Lie algebra of $E_2^c$).
These three representations are distinguished by the number of ZGA
highest/lowest weight
vectors they contain and by the eigenvalue, $t$, of the central element.
Type I representations contain neither ZGA highest nor lowest weight
vectors and are characterized by $t=0$. Type II representations contain
a ZGA lowest weight state and are characterized by $t>0$, while in Type III
representations
there is a ZGA highest weight state and all the states have $t<0$.

The spectrum and some
scattering processes of the string theory obtained by adding $22$
flat Euclidean coordinates to the $E_2^c$ WZW model have been investigated
in [5]. Here we analyze this string theory,
that we call Nappi Witten string,
from a different point of
view. The problem is that
all the KM representations based on Type I-III ZGA representations
are plagued by the presence of
negative norm states. These negative norm states then
appear in the string state space and one may ask whether the Virasoro
conditions are enough to remove them from the physical subspace.
There are at least two examples where the Virasoso conditions do indeed
guarantee -- albeit in a slightly different way -- that the physical
subspace contains only states with non-negative norms: the ordinary
bosonic string propagating on a flat $D=26$ dimensional Minkowski
space [6] and the example of the $SU(1,1)$ string [7].  The difference
between these two cases lies in the fact that for the $SU(1,1)$ string one
has to truncate the string state space to contain only a restricted set
of $\widehat {SU}(1,1)$ representations [7], while for the bosonic string on
Minkowski space no such truncation is necessary.

Recently, following the method outlined in [8] for the bosonic string
on Minkowski space, we proved an analogous theorem [9] for the Nappi Witten
string as defined above. This theorem states, that the Virasoro conditions
are indeed enough to guarantee that the states in the physical subspace
have non negative norms at least when we use KM representations having
Type II states at the base and $1-t/(2k)>0$ or when
we use KM representations
with Type III base and $1+t/(2k)>0$. (Here $k$ denotes the parameter of the
WZW model, that appears also as the level of the KM algebra).

Although the proof of our theorem makes it possible to determine the global
$E_2^c\times E_2^c$ quantum numbers of the physical states
 it offers very little insight into the structure of the physical
subspace. Therefore, in the present paper, we reconsider the quantization of
the Nappi Witten string in the light-cone gauge
using the old fashioned operator approach.
The motivation to consider
the light-cone quantization comes from various considerations:

\item{$\bullet$} This way one works with positive norm states only,
thus the spectra one finds are the physical ones.

\item{$\bullet$} One can  confirm the spectra obtained by
 covariant methods and compare the two methods on an example other than
that of a string propagating on flat Minkowski space.

\item{$\bullet$} The light-cone quantization may give new
insight into the interpretation of the various peculiar states of the
Nappi-Witten string.

\item{$\bullet$} One may obtain a physical interpretation of the
restriction on the allowed $\widehat E_2^c$ representations (i.e. the
$t$ range) found in our theorem.

\item{$\bullet$} The direct determination of the physical states makes
the analysis of modular invariance readily accessible.

The viability of light cone quantization in this case is based on an
observation of Horowitz and Steif [2]: the light-cone gauge can be imposed
in a curved spacetime only if it admits a covariantly constant null
vector; i.e. if it describes a plane fronted gravitational wave (an
axion field may also be present).

The paper is organized as follows: in chapter 1. we describe the action and
the classical Hamiltonian in the light-cone gauge. Chapter 2. contains the
discussion of the global symmetries. The mode expansion and quantization is
carried out in chapter 3, while the mass operator, the energy spectrum and
the restrictions on the physical states are analized in chapter 4. In
chapter
5. we discuss the problems related to the determination of the critical
dimension and describe the details of the chiral algebra. Chapter 6. is
devoted
to a brief study of the consequences of dropping the previous limitation
on
the total momentum of the center of mass. Finally, in chapter 7.
the one loop partition function is derived and its modular properties are
analyzed. We make our conclusions in chapter 8.

\centerline{\underbar{ 1. The action and Hamiltonian in the light-cone
gauge}}
\bigskip

As a preparation we transform the spacetime metric
and antisymmetric tensor field
of the Nappi-Witten solution
$$
ds^2=(da^i)^2-2dU(dV+{1\over 2}\epsilon_{ij}a^ida^j)+bdU^2;\qquad
b_{ij}=U\epsilon_{ij},\eqno(1)
$$
($i,j=1,2$) into a form
for which the vanishing of the anomaly has been shown at all loop orders,
and which is also more convenient for the light-cone quantization. Indeed
making the gauge transformation $b_{\mu\nu}\rightarrow b_{\mu\nu}+
2\partial_{[\mu}\lambda_{\nu]}$ ($\mu =U,V,i$) with $\lambda_i={1\over 2}
U\epsilon_{ij}a^j$ on the antisymmetric tensor field and
rotating the $a^i$ fields by an angle $U/2$:
$$
\eqalign{
x^1&=a^1{\rm cos}{U\over 2}+a^2{\rm sin}{U\over 2}\cr
x^2&=-a^1{\rm sin}{U\over 2}+a^2{\rm cos}{U\over 2}\cr}
\eqno(2)
$$
we get
$$
ds^2=(dx^i)^2-2dUdV+(b-(x^i)^2/4)dU^2;\qquad b_{iu}=-(1/2)\epsilon_{ij}x^j.
\eqno(3)
$$
This expression shows that the Nappi-Witten background indeed describes
a plane fronted gravitational wave with rather simple (constant) polarization.

To define the action of the Nappi-Witten string properly
we add extra coordinates,
$x^A$, describing $d$ flat Euclidean dimensions,
to the four ones already present in eq.(3):
$$\eqalign{
S=&{1\over 2\pi}\int d\tau d\sigma\bigl[\partial_ax^A\partial^ax^A+
{k\over2}\bigl(
\partial_ax^i\partial^ax^i-2\partial_aU\partial^aV+(b-(x^i)^2/4)\partial_aU
\partial^aU\cr
&-\epsilon_{ij}x^j\partial_ax^i\partial_bU\epsilon^{ab}\bigr)\bigr].\cr}
\eqno(4)
$$
In eq.(4) we fixed the conformal gauge on the world sheet:
$h_{ab}=\eta_{ab}$,
($a=\tau ,\sigma$), with $\eta_{\tau\tau}=-\eta_{\sigma\sigma}=1$ and also
set the the slope parameter, $\alpha^\prime=1/2$, in spite of considering
the case of closed rather than open strings. In the $x^i$, $U$, $V$ part
of this action we wrote explicitly  the $k$ parameter of the original Nappi
Witten action.

The equations of motion following from eq.(4) are
$$\eqalign{
(\pt^2-\ps^2)x^A=&0,\cr
(\pt^2-\ps^2)U=&0,\cr
2(\pt^2-\ps^2)x^i+{x^i\over2}\partial_aU\partial^aU+2\epsilon^{ji}&(\pt
x^j\ps U-\ps x^j\pt U)=0,\cr
-2(\pt^2-\ps^2)V+2\partial_a\bigl[(b-{(x^i)^2\over4})\partial^aU\bigr]&+
2\epsilon^{ij}\pt x^j\ps x^i=0.\cr}
\eqno(5)
$$
These equations are supplemented by the vanishing of the world sheet
energy momentum tensor, $T_{ab}$, expressing the reparametrization invariance
of the string action.

The light-cone gauge on the target space
is fixed by setting $U=P\tau$, which obviously solves the free equation
for $U$
in (5). The equations of motion for
the $x^i$ coordinates simplify in this gauge:
$$
2(\pt^2-\ps^2)x^i+{x^i\over2}P^2+2P\epsilon^{ij}\ps x^j=0,
\eqno(6)
$$
while the $x^A$ still has to satisfy
the free equations appearing in (5). The two independent
components of the constraint $T_{ab}=0$ take the form
$$
\eqalign{
P\pt V=&{1\over 2}\bigl[ (\pt x^i)^2+(\ps x^i)^2+
P^2(b-{(x^i)^2\over 4})+{2\over k}\bigl( (\pt x^A)^2+(\ps x^A)^2
\bigr)\bigr],\cr
P\ps V=&\pt x^i\ps x^i+{2\over k}\pt x^A\ps x^A.\cr}
\eqno(7)
$$
If $P\ne 0$
we use the two equations appearing in (7) to determine the second light-cone
 coordinate, $V$, in terms of $x^i$ and $x^A$. Indeed one readily
verifies, that $V$ determined this way satisfies not just the
integrability condition, but also the equation of motion in (5) whenever
$x^i$ ($x^A$) satisfy their equation of motion, eq.(6) (eq.(5) respectively).
Since all fields are periodic in the case of closed strings, integrating
the second equation in (7) over $\sigma$ we get the usual constraint
$$
\int\limits_0^{2\pi} d\sigma (\pt x^i\ps x^i+{2\over k}
\pt x^A\ps x^A)=0,\eqno(8)
$$
which enforces invariance under shifting $\sigma$ by a constant.

{}From the light-cone form of the action
$$\eqalign{
S=&{1\over 2\pi}\int d\tau d\sigma\bigl[\partial_ax^A\partial^ax^A+
{k\over2}\bigl(
\partial_ax^i\partial^ax^i-2P\pt V+(b-(x^i)^2/4)P^2\cr
&+P\epsilon_{ij}x^j\ps x^i\bigr)\bigr]\cr}
\eqno(9)
$$
one readily derives the canonical momenta
$$
\Pi^i={1\over\pi}{k\over2}\pt x^i,\qquad \Pi^A={1\over\pi}\pt x^A,
\eqno(10)
$$
and constructs the canonical Hamiltonian in the usual way:
$$\eqalign{
H=&\int\limits_0^{2\pi}{d\sigma\over 2\pi}\bigl[(\pi)^2(\Pi^A)^2+(\ps x^A)^2+
{2\over k}(\pi)^2(\Pi^i)^2\cr
+{k\over2}&\bigl((\ps x^i)^2-(b-(x^i)^2/4)P^2
-P\epsilon_{ij}x^j\ps x^i\bigr)\bigr]\cr
=&\int\limits_0^{2\pi}{d\sigma\over 2\pi}\bigl[(\pt x^A)^2+(\ps x^A)^2+
{k\over2}\bigl(
(\pt x^i)^2+(\ps x^i)^2-(b-(x^i)^2/4)P^2\cr
&-P\epsilon_{ij}x^j\ps x^i\bigr)\bigr].\cr}
\eqno(11)
$$
\vfill
\eject
\centerline{\underbar{2. Global symmetries}}
\bigskip

The symmetries of the target space appear in string theory as global (world
sheet independent) internal symmetries. Since they play an essential role
in the quantization and in the interpretation of the results,
we describe them in some details. For the theory described by eq.(4)
there are three types of global symmetries:
the ones related to the four \lq\lq interesting" coordinates of the Nappi
Witten solution, the ones belonging to the $x^A$ part of the action and the
ones mixing these two types of coordinates.
We are mainly interested in the first and third types of these global
symmetries  since the symmetries of the $x^A$ part form the
standard $E_d$. A particularly interesting question is
the possible reduction of
these symmetries when one moves to the light-cone gauge; i.e. when one
replaces eq.(4) by eq.(9).

As for the first type of global symmetries we note that
the Lagrangian in eq.(4) is obviously invariant under the (infinitesimal)
translation of the $U$ or the $V$ coordinates:
$$\eqalign{
T_{u_0}:&\quad x^i\rightarrow x^i;\quad U\rightarrow U+u_0;
\quad V\rightarrow V,\cr
T_{v_0}:&\quad x^i\rightarrow x^i;\quad U\rightarrow U;\quad
V\rightarrow V+v_0,\cr}
$$
while it changes by a total divergence under the rotation of the $x^i$
coordinates:
$$
T_\alpha:\quad x^i\rightarrow x^i+\alpha \epsilon^{ij}x_j;\quad
U\rightarrow U;\quad V\rightarrow V.
$$
A direct computation shows that the action in eq.(4) is also invariant under
the four \lq\lq twisted translations" of the $x^i$ coordinates:
$$\eqalign{
T_{\delta_1}&:\quad x_1\rightarrow x_1+\delta_1{\rm cos}(U/2);
\quad x_2\rightarrow x_2-\delta_1{\rm sin}(U/2);\quad
 U\rightarrow U;\cr
 & V\rightarrow V-(\delta_1/2)(x_1{\rm sin}(U/2)+x_2{\rm cos}(U/2)),\cr
T_{\delta_2}&:\quad x_1\rightarrow x_1+\delta_2{\rm sin}(U/2);
\quad x_2\rightarrow x_2+\delta_2{\rm cos}(U/2);\quad
 U\rightarrow U;\cr
 & V\rightarrow V-(\delta_2/2)(x_2{\rm sin}(U/2)-x_1{\rm cos}(U/2)),\cr
T_{\Delta_1}&:\quad x_1\rightarrow x_1+\Delta_1{\rm cos}(U/2);
\quad x_2\rightarrow x_2+\Delta_1{\rm sin}(U/2);\quad
 U\rightarrow U;\cr
 & V\rightarrow V-(\Delta_1/2)(x_1{\rm sin}(U/2)-x_2{\rm cos}(U/2)),\cr
T_{\Delta_2}&:\quad x_1\rightarrow x_1-\Delta_2{\rm sin}(U/2);
\quad x_2\rightarrow x_2+\Delta_2{\rm cos}(U/2);\quad
 U\rightarrow U;\cr
 & V\rightarrow V-(\Delta_2/2)(x_2{\rm sin}(U/2)+x_1{\rm cos}(U/2)).\cr}
$$
Using these infinitesimal transformations one can compute the Lie algebra
of the 7 parameter symmetry group, $M_7$. We find that the two types of
translations commute with each other, while among themselves they yield
the $V$ translation:
$$
[T_{\delta_i},T_{\Delta_j}]=0,\quad [T_{\delta_2},T_{\delta_1}]=
-[T_{\Delta_2},T_{\Delta_1}]=T_{v_0}.
\eqno(12)
$$
Both the rotation and the $U$ translation act on the $x^i$ translations:
$$
[T_{\alpha},T_{\delta_i}]=-\epsilon_{ij}T_{\delta_j},\quad
[T_{u_0},T_{\delta_i}]={1\over2}\epsilon_{ij}T_{\delta_j},
\eqno(13)
$$
$$
[T_{\alpha},T_{\Delta_i}]=-\epsilon_{ij}T_{\Delta_j},\quad
[T_{u_0},T_{\Delta_i}]=-{1\over2}\epsilon_{ij}T_{\Delta_j}.
\eqno(14)
$$
It is easy to see from eq.(13-14) that one can form two independent
combinations of the rotation and $U$ translation, ($T_{u_0}\pm T_\alpha /2$),
that act only on one
type of the $x^i$ translations, leaving the other type invariant:
$$
\eqalign{[T_{u_0}+ T_\alpha /2,T_{\delta_i}]=&0,\quad
[T_{u_0}+ T_\alpha /2,T_{\Delta_i}]=-\epsilon_{ij}T_{\Delta_j},\cr
[T_{u_0}- T_\alpha /2,T_{\Delta_i}]=&0,\quad
[T_{u_0}- T_\alpha /2,T_{\delta_i}]=\epsilon_{ij}T_{\delta_j}.\cr}
$$
 These
linear combinations, together with the \lq\lq affected" translations and
the $V$ translation form two (chiral) copies of the $E_2^c$ algebra
inherent in
the Nappi-Witten construction\footnote{$^\dagger$}
{C. Duval [unpublished] has shown that this seven parameter
symmetry is indeed the full isometry of the metric (3).}.

The conserved currents belonging to these symmetries are derived in the
usual way; we list only some of them:
$$\eqalign{
J_a^{\delta_1}&={k\over4\pi}\bigl[
(2\partial_ax^1-x^2\partial_bU\epsilon^{ab}){\rm cos}(
{U\over2})-(2\partial_ax^2+x^1\partial_bU\epsilon^{ab}){\rm sin}(
{U\over2})\cr
&+\partial_aU((x_1{\rm sin}({U\over2})+x_2{\rm cos}({U\over2}))\bigr],\cr
J_a^{v_0}&=-{k\over2\pi}\partial_aU,\cr
J_a^{u_0}&={k\over4\pi}\bigl[-2\partial_aV+2(b-{(x^i)^2\over4})\partial_aU
-\epsilon^{ij}\epsilon^{ba}x^j\partial_bx^i\bigr],\cr
J_a^{\alpha}&={k\over4\pi}\bigl[2\partial_ax^i-\epsilon^{ij}x_j\partial_bU
\epsilon^{ab}\bigr]\epsilon^{im}x_m.\cr}
\eqno(15)
$$
Integrating the $\tau$
components of these currents over $\sigma$ we get the conserved charges
belonging to the various symmetries.
 It is interesting to notice, that
the currents belonging to the various $x^i$ translations
are linear in $x^i$, and it is only the current of the $U$ translation that
depends on the second light-cone coordinate, $V$.

Comparing the light cone form of the
action, eq.(9), and the covariant one in eq.(4), we see that the
transformations generating the first type of global symmetries of eq.(4)
 -- with the exception
of the $U$ translations -- are also explicit symmetries of eq.(9). This
situation is significantly different from the case of strings propagating
on flat Minkowski space, when going to the light cone gauge reduces the
explicit $SO(D-1,1)\triangleleft T^D$ symmetry of the covariant action to
a manifest $SO(D-2)\triangleleft T^{D-2}$ only.

In the light-cone gauge, where $\pt U=P$, $\ps U=0$, eq.(15)
simplifies slightly. We find
trivially $Q^{v_0}=-kP$, while a short computation using eq.(7) and eq.(11)
yields:
$$
Q^{u_0}\equiv \int\limits_0^{2\pi}d\sigma J_{\tau}^{u_0}=-{1\over P}H.
\eqno(16)
$$
This shows that all seven charges of the covariant action, that belong to
the first type of global symmetries,
are also
conserved in the light cone gauge.

In the quantum theory (or in the classical one using the Poisson brackets)
these charges satisfy an algebra similar to the one given by eq.(12-14).
This algebra has three independent Casimir operators, that can be chosen in
the following way: a linear one $C_1=Q^{v_0}$, and two quadratic ones
$$\eqalign{
C_2&=4Q^{u_0}Q^{v_0}-(Q^{\delta_i})^2-(Q^{\Delta_i})^2,\cr
C_3&=2Q^{\alpha}Q^{v_0}+(Q^{\delta_i})^2-(Q^{\Delta_i})^2.\cr}
\eqno(17)
$$
 In terms of the
\underbar{chiral} $E_2^c$ algebra $C_2$ correponds to
 the sum while $C_3$ to the
difference of the left and right Casimirs.

The symmetries of the covariant action mixing the \lq\lq Nappi Witten"
and the $x^A$ coordinates
are generated by the infinitesimal transformations:
$$
T^\beta_A:\quad
x^A\rightarrow x^A+{k\over2}\beta U,\quad x^B\rightarrow x^B\ B\ne A,\quad
U\rightarrow U,\quad V\rightarrow V+\beta x^A.
$$
(Here $\beta$ is an infinitesimal parameter). Using the infinitesimal
transformations of the $E_d$ symmetry:
$$
\eqalign{
T_A&:\quad x^A\rightarrow x^A+\delta ^A,\quad U\rightarrow U,
\quad V\rightarrow V,\cr
T_{AB}&:\quad x^C\rightarrow x^C+\gamma (\delta^{AC}x^B-\delta^{BC}x^A),
\quad U\rightarrow U,
\quad V\rightarrow V,\cr}
$$
one readily proves that
$$
[T^\beta_A,T_{u_0}]={k\over2}T_A,\quad [T^\beta_A,T_B]=\delta_{AB}T_{v_0},
\quad
[T_{AB},T^\beta_C]=\delta_{BC}T^\beta_A-\delta_{AC}T^\beta_B.
$$
Furthermore it is easy to see that the symmetries generated by $T^\beta_A$
are also explicit
symmetries of the light cone action, eq.(9).
The conserved current belonging to the $T^\beta_A$ transformation
has the form:
$$
J^\beta_a={k\over2\pi}(\partial_ax^AU-x^A\partial_aU)
$$
thus in the light cone gauge the conserved charge associated to it becomes
$$
Q^\beta_A=-{kP\over2\pi}\int\limits_0^{2\pi}d\sigma (\tau\pt x^A-x^A).
$$
\bigskip

\centerline{\underbar{3. Oscillator expansion and quantization}}
\bigskip

In the light-cone gauge the equation of motion for the $x^i$ fields,
eq.(6), is not only linear in the $x^i$-s but contains only constant
coefficients. Therefore decomposing the $x^i$ fields into Fourier
modes
$$
x^i(\tau,\sigma)=\sum\limits_nX^i_n(\tau)e^{in\sigma}
\eqno(18)
$$
we find that the Fourier coefficients, $X^i_n$ ($i=1,2$), satisfy  coupled
oscillator type equations for all $n$:
$$\eqalign{
2{d^2X^1_n\over d\tau^2}+(2n^2+{P^2\over 2})X^1_n+2PinX^2_n&=0,\cr
2{d^2X^2_n\over d\tau^2}+(2n^2+{P^2\over 2})X^2_n-2PinX^1_n&=0.\cr}
\eqno(19)
$$
Making the ansatz $X^i_n=C^i_ne^{i\omega_n\tau}$ we find
$$
\omega_n^2=(n\pm {P\over 2})^2.
\eqno(20)
$$
Either directly from eq.(6), or from eq.(19-20), we see that the \lq\lq zero
modes" ($n=0$), that describe the motion of the string's center of mass,
also satisfy  oscillator equations. This behaviour is strikingly different
from the usual, uniform motion of the CM coordinate in flat dimensions.
The frequency of these CM oscillations
is $P/2$; this precisely corresponds to the coefficient of the quadratic
potential in eq.(9).

The interesting property of the $\omega_n=\pm\vert n\pm P/2\vert$
eigenfrequencies (we need all
four combinations!) is that substituting them into eq.(18) we get \lq almost'
chiral expressions. Indeed, when $0<P/2<1$, we find --
having carefully determined
the eigen-directions of the matrix originating from eq.(19) -- that:
$$\eqalign{
x^1+ix^2&={1\over\sqrt{k}}\Bigl[{2\over P}K^\dagger e^{i\Pt}+
{2\over P}\Lambda e^{-i\Pt}+e^{i\Pt}\sum\limits_{n>0}\bigl(
{\lambda^+_n\over n-P/2}e^{-in(\tau+\sigma)}\cr
&+{(k^+_n)^\dagger\over n+P/2}e^{in(\tau+\sigma)}\bigr)+
e^{-i\Pt}\sum\limits_{n>0}\bigl(
{\lambda^-_n\over n+P/2}e^{-in(\tau-\sigma)}\cr
&+{(k^-_n)^\dagger\over n-P/2}e^{in(\tau-\sigma)}\bigr)\Bigr]\cr}
\eqno(21)
$$
and
$$\eqalign{
x^1-ix^2&={1\over\sqrt{k}}\Bigl[{2\over P}Ke^{-i\Pt}+
{2\over P}(\Lambda)^\dagger e^{i\Pt}+e^{-i\Pt}\sum\limits_{n>0}\bigl(
{(\lambda^+_n)^\dagger\over n-P/2}e^{in(\tau+\sigma)}\cr
&+{k^+_n\over n+P/2}e^{-in(\tau+\sigma)}\bigr)+
e^{i\Pt}\sum\limits_{n>0}\bigl(
{(\lambda^-_n)^\dagger\over n+P/2}e^{in(\tau-\sigma)}\cr
&+{k^-_n\over n-P/2}e^{-in(\tau-\sigma)}\bigr)\Bigr] .\cr}
\eqno(22)
$$
Here the various normalizations are introduced for later convenience and for
the classical Fourier coefficients $\dagger$ stands for complex conjugation.
The distribution of complex conjugated quantities in eq.(21-22) guaranties
that $(x^1+ix^2)^\dagger=x^1-ix^2$.

We quantize our $x^i$ fields by promoting the various Fourier components
appearing in eq.(21-22) to operators. We fix their commutators
by requiring, that the canonical equal time commutation relations
$$\eqalign{
[\Pi^i(\tau,\sigma),x^j(\tau,\sigma^\prime)]&=-i\delta^{ij}\delta(\sigma-
\sigma^\prime),\cr
[\Pi^i(\tau,\sigma),\Pi^j(\tau,\sigma^\prime)]&=0,
\quad [x^i(\tau,\sigma),x^j(\tau,\sigma^\prime)]=0,\cr}
$$
hold when $\Pi^i(\tau,\sigma)$ is expressed in terms of $x^i$, as in eq.(10).
Proceeding this way we find that the only nonvanishing commutators for the
non zero modes (i.e. $n,m>0$)
are the
following:
$$\eqalign{
[k^+_n,(k^+_m)^\dagger]=&\delta_{n,m}(n+P/2),\quad
[\lambda^+_n,(\lambda^+_m)^\dagger]=\delta_{n,m}(n-P/2),\cr
[k^-_n,(k^-_m)^\dagger]=&\delta_{n,m}(n-P/2),\quad
[\lambda^-_n,(\lambda^-_m)^\dagger]=\delta_{n,m}(n+P/2),\cr}
\eqno(23)
$$
while for the zero mode operators
$$
[\Lambda,(\Lambda)^\dagger]=P/2;\quad [K,(K)^\dagger]=P/2
\eqno(24)
$$
results. Eq.(23) shows, that in both light-cone directions we have a
set of $n+P/2$ modded and a set of $n-P/2$ modded oscillators. Eq.(24)
implies that the oscillatory nature of the CM coordinate's motion persists
to the quantum theory.

The appearance of these non-integrally modded (\lq\lq twisted")
oscillators can be understood in the following way: Rewriting
eq.(6) as:
$$
\dal (x^1\pm ix^2)+{P^2\over4}(x^1\pm ix^2)\mp iP\ps (x^1\pm ix^2)=0,
$$
and introducing the complex $\Phi (\tau,\sigma)$ field through the
ansatz:
$$
x^1+ix^2=e^{-i{P\over2}\sigma}\Phi (\tau,\sigma);\qquad
x^1-ix^2=e^{i{P\over2}\sigma}\Phi^\dagger (\tau,\sigma);
$$
one easily deduces that $\Phi$ obeys the free field equation:
$\dal \Phi =0$. Thus $x^1\pm ix^2$ can be written as:
$$
x^1+ix^2=e^{-i{P\over2}\sigma}\bigl(\phi (\tau-\sigma)+
\psi^\dagger (\tau+\sigma)
\bigr);\quad
x^1-ix^2=e^{i{P\over2}\sigma}\bigl(\phi^\dagger (\tau-\sigma)+
\psi (\tau+\sigma)\bigr).
$$
However, because of the $\exp (\pm iP\sigma/2)$ prefactor, the free
fields, $\phi (\tau-\sigma)$ and $\psi (\tau+\sigma)$, must be
\lq\lq twisted" (i.e. must satisfy non-periodic boundary conditions
in $\sigma$) to yield periodic $x^1\pm ix^2$-s.

Since the equations of motion for the $x^A$ coordinates are the standard
ones we get the standard oscillator expansion for them:
$$
x^A=q^A+{p^A\over2}\tau+{i\over2}\sum\limits_{n\ne 0}\bigl(
{\alpha^{-A}_n\over n}e^{-in(\tau-\sigma)}+
{\alpha^{+A}_n\over n}e^{-in(\tau+\sigma)}\bigr).
\eqno(25)
$$

\centerline{\underbar{4. Hamilton operator, $M^2$ and physical states}}
\bigskip

Using the oscillator expansions eq.(21,22,25) in eq.(11) we find the
Hamilton operator:
$$\eqalign{
H&={1\over4}(p^A)^2+\sum\limits_{n>0}(\alpha^{-A}_{-n}\alpha^{-A}_n+
\alpha^{+A}_{-n}\alpha^{+A}_n)\cr
&+\sum\limits_{n>0}((\lambda^{+}_{n})^\dagger\lambda^{+}_n+
(\lambda^{-}_{n})^\dagger\lambda^{-}_n+
(k^{+}_{n})^\dagger k^{+}_n+(k^{-}_{n})^\dagger k^{-}_n)\cr
&+(K)^\dagger K+(\Lambda)^\dagger\Lambda -{k\over2}bP^2+a.\cr}
\eqno(26)
$$
Here $a$ is a normal ordering constant, that can be determined by $\zeta$
function regularization since it originates in reordering the integer
modded oscillators, the zero mode ones and the $n\pm P/2$ modded ones:
$$\eqalign{
a&=2d{1\over2}(-{1\over12})+2{1\over2}{P\over2}+2{1\over2}(
\sum\limits_{n=1}^{\infty}(n+P/2)+\sum\limits_{n=1}^{\infty}(n-P/2))\cr
&=(-{d\over12})+{P\over2}+\zeta(-1,P/2)+1-P/2+\zeta(-1,1-P/2)\cr
&=(-{d+2\over12})+1+{P\over2}(1-{P\over2}).\cr}
\eqno(27)
$$
{}From eq.(26-27) one can read off the spectrum, but to analyse it we
have to determine what are the restrictions on the physical states. Also
to put the spectrum of the Hamiltonian into an appropriate perspective
we have to derive the analog of the $M^2$ operator for the gravitational
background described in eq.(4).

In the usual case, when strings propagate on a flat Minkowski background,
the $M^2$ operator is nothing but the quadratic Casimir of the symmetry
group constructed of the generators of translations. Therefore in the present
case the appropriate generalisation is to consider a linear combination
of $C_2$ and $(p^A)^2$. The relative coefficient between them
is determined from
requiring it to reproduce the well known result in the limiting case of
flat Minkowski space:
$$
M^2={1\over k}\bigl(4Q^{u_0}Q^{v_0}-(Q^{\delta_i})^2-(Q^{\Delta_i})^2
\bigr)-p^Ap^A.
\eqno(28)
$$
Since the charges $Q^{\delta_i}$, $Q^{\Delta_i}$, contain only the zero
mode operators (e.g.
$
Q^{\delta_1}=i\sqrt{k}(\Lambda^\dagger-\Lambda)$, $
Q^{\delta_2}=-\sqrt{k}(\Lambda^\dagger+\Lambda)$),
when expressed in terms of the oscillators, $M^2$ has the form:
$$\eqalign{
M^2&=4\Bigl[\sum\limits_{n>0}(\alpha^{-A}_{-n}\alpha^{-A}_n+
\alpha^{+A}_{-n}\alpha^{+A}_n)\cr
&+\sum\limits_{n>0}((\lambda^{+}_{n})^\dagger\lambda^{+}_n+
(\lambda^{-}_{n})^\dagger\lambda^{-}_n+
(k^{+}_{n})^\dagger k^{+}_n+(k^{-}_{n})^\dagger k^{-}_n)\cr
&-{k\over2}bP^2+a-P/2\Bigr] .\cr}
\eqno(29)
$$
This expression also shows, that the zero mode operators, that are present in
$H$, disappear from $M^2$.

The physical states are restricted by the left-right symmetry condition:
i.e. eq.(8) imposes
on them the invariance under shifting the $\sigma$ coordinate.
Carrying out the integration in eq.(8) we get:
$$\eqalign{
&\sum\limits_{n>0}\bigl[\alpha^{-A}_{-n}\alpha^{-A}_n+
{n\over n+P/2}(\lambda^{-}_{n})^\dagger\lambda^{-}_n
+{n\over n-P/2}(k^{-}_{n})^\dagger k^{-}_n\bigr]\cr
&-\sum\limits_{n>0}\bigl[\alpha^{+A}_{-n}\alpha^{+A}_n
+{n\over n-P/2}(\lambda^{+}_{n})^\dagger\lambda^{+}_n+
{n\over n+P/2}(k^{+}_{n})^\dagger k^{+}_n\bigr]=0.\cr}
\eqno(30)
$$
Two things must be remarked about this equation that generalizes the usual
left right symmetry condition:  it requires that the
total excitation numbers of the \lq\lq $n$ index" oscillators be the same
for the $+$ and $-$ light cone directions, irrespective of how these numbers
are distributed among the $n$, $n+P/2$ and $n-P/2$ modded ones. In the
same time eq.(30) says nothing about the zero mode oscillators.
The absence of the zero mode operators
 in $M^2$ and in eq.(30) implies, that physical states with the same
$M^2$ are infinitely degenerate. Indeed  denoting the excitation
numbers of the $\Lambda$ and $K$ oscillators
by $n_\Lambda$ and $n_K$ respectively and
 making the replacements
$$\tilde n_\Lambda=
n_\Lambda+N,\qquad\tilde n_K=n_K+M\quad (N,M>0)
\eqno(31)$$
generates new physical states from
the old ones. These new states correspond to placing the
 string as a whole -- no matter how it is \lq\lq wiggling" --
on the various discrete energy levels in the
external harmonic oscillator potential.

Using eq.(20,21, and 25) in eq.(15) to compute $Q^\alpha$ we get:
$$\eqalign{
Q^\alpha={2\over P}(K^\dagger K-\Lambda^\dagger\Lambda)&-
\sum\limits_{n>0}\bigl[{1\over n+P/2}[(k^{+}_{n})^\dagger k^{+}_n
-(\lambda^{-}_{n})^\dagger\lambda^{-}_n]\cr
&+{1\over n-P/2}[(k^{-}_{n})^\dagger k^{-}_n
-(\lambda^{+}_{n})^\dagger\lambda^{+}_n]
\bigr].\cr}
\eqno(32)
$$
Using this expression one can write
the new left right symmetry condition, eq.(30), in an alternative
form:
$$\eqalign{
&\sum\limits_{n>0}\bigl[\alpha^{-A}_{-n}\alpha^{-A}_n+
(\lambda^{-}_{n})^\dagger\lambda^{-}_n
+(k^{-}_{n})^\dagger k^{-}_n\bigr]+K^\dagger K\cr
&-\sum\limits_{n>0}\bigl[\alpha^{+A}_{-n}\alpha^{+A}_n
+(\lambda^{+}_{n})^\dagger\lambda^{+}_n+
(k^{+}_{n})^\dagger k^{+}_n\bigr]-\Lambda^\dagger\Lambda
-{P\over2}Q^\alpha=0.\cr}
\eqno(33)
$$

\centerline{\underbar{5. Critical dimension and the chiral algebra}}
\bigskip

The next step is to try to derive the critical dimension and the intercept
(i.e. the normal ordering constant $a$)
in the light cone gauge. In the case of strings on
flat Minkowski space these numbers are obtained by requiring the complete
$SO(D-1,1)$
symmetry algebra of the covariant action to hold even in the light cone
formalism where only its $SO(D-2)$ part is a manifest symmetry.
As we mentioned earlier for the
Nappi Witten string it is only the $U$ translation that is not an
explicit symmetry of the light cone action, eq.(9). Therefore the only
thing we can impose is the vanishing of the commutators between $Q^{u_0}$
and those charges that depend non-linearly on the oscillators, i.e. the
vanishing of
$[Q^{u_0},Q^\alpha ]$
and $[Q^{u_0},M^{AB}]$, where
$$
M^{AB}={1\over2\pi}\int\limits_0^{2\pi}d\sigma (x^A\pt x^B-x^B\pt x^A).
$$
However a simple computation, using the explicit form of these charges,
shows that these commutators vanish without imposing any restriction
either on $d$ or on $a$. One can also check that all commutators of the
charges $Q^\beta_A=-kPq^a$, belonging to the symmetries mixing the \lq\lq
 Nappi Witten" coordinates and the $x^A$ ones, are the required ones
independently of the values of $d$ and $a$. Therefore the critical dimension
is not determined in this light cone quantization of the Nappi Witten
string.
Though surprising, this conclusion is in accord with what was found
in ref.[2] among somewhat similar circumstances.\footnote{$^\dagger$}{The
critical dimension in the
light-cone quantization of $2D$ sigma models was derived recently in
ref.[11],
using the functional integral formalism.}

It is also interesting to determine how the $\hat E_2^c$ chiral
Kac Moody algebra
of the Nappi Witten construction gets distorted in the light cone gauge.
In the covariant formalism the components of the holomorphic current,
$\partial gg^{-1}$, ($\partial=\pt -\ps$)
that belong to the various generators of $E_2^c$
in the $g=\exp [a_1P_1+a_2P_2]\exp [UJ-VT]$ parametrization, are
given by [3,4]:
$$
\eqalign{
P_1(\xi_-)\equiv J^1(\xi_-)=&k(\partial a_1+a_2\partial U)\cr
P_2(\xi_-)\equiv J^2(\xi_-)=&k(\partial a_2-a_1\partial U)\cr
J(\xi_-)\equiv J^3(\xi_-)=&k(-\partial V+{1\over2}\epsilon^{ij}
a_i\partial a_j+(b-{1\over2}a_ia_i)\partial U)\cr
T(\xi_-)\equiv J^4(\xi_-)=&k\partial U.\cr}
\eqno(34)
$$
Here we introduced $\xi_-=\tau-\sigma$ and also define the $J^k_n$
Fourier coefficients
of $J^k(\xi_-)$ $k=1,..4$ as
$$
J^k(\xi_-)=\sum\limits_{-\infty}^\infty J^k_ne^{-in\xi_-}.
$$
Using the inverse of eq.(2) it is easy to express $J^k$ in terms of our
$x^i$ fields. The fact that
in the light cone gauge the Kac Moody algebra generated by $J^k_n$ gets
distorted can be seen already from the defining relation,
$U=P\tau$, since this effectively sets
$J^4_n=kP\delta_{n,0}$. Nevertheless  a straightforward computation
using the inverse of eq.(2) and the
oscillator expansions in eq.(21,22) yields ($n>0$):
$$
\eqalign{
J^1_0=i\sqrt{k}(\Lambda^\dagger-\Lambda)\quad
J^1_n=&-i\sqrt{k}(\lambda^-_n+k^-_n)\quad
J^1_{-n}=i\sqrt{k}((\lambda^-_n)^\dagger+(k^-_n)^\dagger)\cr
J^2_0=-\sqrt{k}(\Lambda^\dagger+\Lambda)\quad
J^2_n=&\sqrt{k}(k^-_n-\lambda^-_n)\quad
J^2_{-n}=\sqrt{k}((k^-_n)^\dagger-(\lambda^-_n)^\dagger).\cr}
\eqno(35)
$$
Now one easily obtaines on account of eq.(22,23) that for any $n,m\in Z$
$$
[J^i_n,J^j_m]=2kn\delta^{ij}_{n+m}+i\epsilon^{ij}kP\delta_{n+m}
\quad i,j=1,2.
\eqno(36)
$$
Eq.(36) shows that the Kac Moody subalgebra generated by $J^1_n$, $J^2_n$
and $J^4_n$ remains unbroken even in the light cone gauge.
The computation of $J^3_n$ -- that exploits the constraints in eq.(7) --
is more involved and we give here only the final result expressed in
terms of the Fourier coefficients of $J^1_n$, $J^2_n$ and $x^A$:
$$
J^3_n=-{1\over P}
\bigl( {1\over2k}(:(J^1)^2:_n+:(J^2)^2:_n)
+\sum\limits_l:\tilde\alpha^{-A}_l
\tilde\alpha^{-A}_{n-l}:+(-{k\over2}bP^2+\hat a)\delta_{n,0}\bigr) ,
\eqno(37)
$$
where $\tilde\alpha^{-A}_0=p^A/2$, $\tilde\alpha^{-A}_k=\alpha^{-A}_k$ if
$k\ne 0$, $::$ denotes the standard normal ordering (i.e. creation operators
stand to the left of annihilation ones) and $\hat a$ is a normal ordering
constant. Eq.(37) leads to the commutators:
$$
[J^3_l,J^i_m]=i\epsilon^{ij}J^j_{l+m}+{2m\over P}J^i_{l+m},
\quad i,j=1,2.\eqno(38)
$$
The appearance of the second term in eq.(38) shows the distortion of
the $\hat E_2^c$ chiral algebra since it would require only the first term.
This modification of the chiral algebra in the light cone gauge is
not a particular property of the Nappi Witten string: it happens
even in the case of strings on flat Minkowski space. Indeed, even in this
case, the commutators between the Fourier coefficients of the chiral currents
belonging to the transverse coordinates, $\partial x^A$, and
the Fourier coefficients of the dependent light cone coordinate, $\partial V$,
(as computed from the vanishing of the world sheet energy momentum tensor),
are entirely analogous to the second term in eq.(38).

We can rephrase the change of the chiral algebra in the light cone gauge
in a slightly different way by introducing the quantities
$$
W_n={1\over4k}
\bigl( :(J^1)^2:_n+:(J^2)^2:_n\bigr) -{P^2\over4}\delta_{n,0}.
\eqno(39)
$$
The commutator
$$
[W_m,W_n]=(m-n)W_{m+n}+{2\over12}m(m^2-1)\delta_{m+n}
$$
shows that one can think of $W_n$ as a sort of light cone Virasoro generator
with central charge $2$. However the analog of eq.(38),
$$
[W_m,J^i_n]=-nJ^i_{n+m}-{i\over2}P\epsilon^{ij}J^j_{n+m},
\quad i,j=1,2,\eqno(40)
$$
means that the currents, $J^i_n$, are not purely weight one primary fields
with respect to this Virasoro algebra.

In the same time we emphasize, that the \lq zero mode algebra' generated by
$J^k_0$, stays undistorted. Furthermore, since $J^4_0=-Q^{v_0}$ and
$J^i_0=Q^{\delta_i}$ for $i=1,2$, this zero mode algebra is identical to
the global, chiral $E_2^c$
one generated by $-Q^{v_0},Q^{\delta_i}$ and $Q^{u_0}-Q^\alpha /2$.
This shows, that the zero mode operators of the string, $\Lambda$ and
$\Lambda^\dagger$, play a multiple role: in addition to describing the
CM motion of the string they also determine the step operators in these
global, chiral $E_2^c$ algebras. Repeating the light-cone analysis with
the antiholomorphic $\hat E_2^c$ current one finds in precisely the same
way that the undistorted algebra of $\bar J^k_0$ is nothing but the one
generated by $-Q^{v_0},Q^{\Delta_i}$ and $Q^{u_0}+Q^\alpha /2$.

These findings have some interesting implications if we want to make
a comparison between the physical states obtained here and in the
covariant formulation. First, one has to
identify $t$, the eigenvalue of the central element, $T_0$,
on physical states with $kP$, the
eigenvalue of $-Q^{v_0}$, and $j_{L,R}$, the eigenvalue of $J_0^{L,R}$,
with the eigenvalue of $-(Q^{u_0}\pm {1\over2}Q^\alpha )=-{1\over P}
(H\mp {P\over2}Q^\alpha)$. Recalling that in the light-cone formalism
the left-right symmetry condition for physical states can be written
in the form of eq.(33), and combining this with the Hamiltonian, eq.(26),
one readily shows that the eigenvalues of $-{1\over P}
(H\mp {P\over2}Q^\alpha)$ on physical states
depend on the excitation numbers of oscillators
belonging to only one of the light-cone directions. Furthermore, taking
the upper sign say and
denoting the excitation numbers of the $k^+_n$, $\lambda^+_n$ and
$\alpha^+_n$ oscillators by $M^+_n$, $L^+_n$ and $C^+_n$ respectively,
as well as introducing $S^+_n=M^+_n+L^+_n+C^+_n$, one finds for the
eigenvalue:
$$
j_+={2\sum\limits_nS^+_n\over P}+\sum (M^+_n-L^+_n)+n_\Lambda +
{p_A^2\over4P}-{1\over2}kbP+{a\over P}.
$$
Here $a$ is the normal ordering constant coming from the Hamiltonian,
and all the oscillator excitation numbers, $M^+_n$, $L^+_n$, $C^+_n$
and $n_\Lambda$ are non negative integers. This $j_+$ describes a
highest weight vector of the aforementioned $E_2^c$ algebra, if
$n_\Lambda=0$. The highest weight values of $j_+$ can be compared to
the ones obtained in the covariant formalism [9],
 with the conclusion, that for $d=22$ (i.e. in the critical
dimension) the two sets of numbers coincide.
\bigskip

\centerline{\underbar{6. Going beyond the $0<P/2<1$ restriction}}
\bigskip

Using the covariant quantization we proved, that in $26$ dimensions
the Nappi Witten string
theory is free of negative norm states iff the continuous parameter,
that characterizes the eigenvalue of the central element of the $E_2^c$
algebra on the physical states, falls into a definite, finite range. The
analog of this parameter in the light cone quantization is the parameter $P$.
Therefore it is interesting to investigate if some sort of restriction on
$P$ arises even in the light cone gauge.

Repeating the procedure leading to eq.(21-24) when $N_0<P/2<N_0+1$ with
$N_0$ being a positive integer we found:
$$\eqalign{
x^1+ix^2&={1\over\sqrt{k}}\Bigl[
e^{i\Pt}\bigl(\sum\limits_{N_0+1}^\infty
{\lambda^+_n\over n-P/2}e^{-in(\tau+\sigma)}
+\sum\limits_{-N_0}^\infty{(k^+_n)^\dagger\over n+P/2}e^{in(\tau+\sigma)}
\bigr)\cr &+
e^{-i\Pt}\bigl(\sum\limits_{-N_0}^\infty
{\lambda^-_n\over n+P/2}e^{-in(\tau-\sigma)}
+\sum\limits_{N_0+1}^\infty{(k^-_n)^\dagger\over n-P/2}e^{in(\tau-\sigma)}
\bigr)\Bigr]\cr}
\eqno(41)
$$
and
$$\eqalign{
x^1-ix^2&={1\over\sqrt{k}}\Bigl[
e^{-i\Pt}\bigl(\sum\limits_{N_0+1}^\infty
{(\lambda^+_n)^\dagger\over n-P/2}e^{in(\tau+\sigma)}
+\sum\limits_{-N_0}^\infty{k^+_n\over n+P/2}e^{-in(\tau+\sigma)}\bigr)\cr &+
e^{i\Pt}\bigl(\sum\limits_{-N_0}^\infty
{(\lambda^-_n)^\dagger\over n+P/2}e^{in(\tau-\sigma)}
+\sum\limits_{N_0+1}^\infty{k^-_n\over n-P/2}e^{-in(\tau-\sigma)}
\bigr)\Bigr] ,\cr}
\eqno(42)
$$
where the Fourier coefficients satisfy an equation, entirely analogous to
eq.(23-24):
$$\eqalign{
[k^+_n,(k^+_m)^\dagger]=&\delta_{n,m}(n+P/2),\quad
[\lambda^-_n,(\lambda^-_m)^\dagger]=\delta_{n,m}(n+P/2),\quad n,m\ge -N_0,\cr
[k^-_n,(k^-_m)^\dagger]=&\delta_{n,m}(n-P/2),\quad
[\lambda^+_n,(\lambda^+_m)^\dagger]=\delta_{n,m}(n-P/2),\quad n,m\ge N_0+1.
\cr}
\eqno(43)
$$
Taking at face value eq.(41-43) reveal only that in both light cone
directions the number of $n+P/2$ modded oscillators \lq\lq exceeds" the
number of $n-P/2$ modded ones but
show no problem with any sort of
normalization for any value of $N_0$. However if one keeps track of the
Fourier coefficients of $e^{\pm i\Pt}e^{\pm in(\tau\pm\sigma)}$ for $0<n<N_0$
as $P$ changes from $0<P/2<1$ to $N_0<P/2<N_0+1$ then one learns, that
the excess
of $n+P/2$ modded oscillators over the $n-P/2$ modded ones is obtained by
changing the interpretation of creation and annihilation operators for those
$n-P/2$ modded oscillators, that, formally, on the basis of eq.(23), would be
\lq\lq wrongly" quantized. As an example consider the $\lambda^+_1$,
$(\lambda^+_1)^\dagger$ pair, the coefficients of $e^{i\Pt}e^{\pm i(\tau +
\sigma)}$ in eq.(21-22),
when $P/2>1$. On the basis of eq.(23) there would be a negative number on the
right hand side of their commutator, however writing $1-P/2$ as $-(-1+P/2)$
and interpreting $-\lambda^+_1$ as
$(k^+_{-1})^\dagger$ and $-(\lambda^+_1)^\dagger$ as $k^+_{-1}$ we get
precisely the form required by eq.(41-43). The necessity of changing creation
and annihilation operators into each other (that corresponds to changing the
ground state) when $P/2$ is increased by one unit is the light cone equivalent
of the presence of negative norm states in the covariant quantization when
the particular parameter is outside of the allowed domain.

\centerline{\underbar{7. One loop partition function, modular properties}}
\bigskip

A bonus of the light-cone quantization in string theory is that it makes
 the computation of the one loop partition function easier, since in this
gauge only the positive norm, physical, transverse states contribute.
If we think of the one loop world sheet, the torus, characterized by a
complex parameter, $\zeta$,
as a cylinder of length $2\pi {\rm Im}\zeta$ whose ends are identified,
then one can
twist the two ends relative to each other by an angle $2\pi {\rm Re}\zeta$
before joining them. The operator which generates this twist is the one that
generates the \lq rigid' translation of the $\sigma$ coordinate, i.e. the one
that appears on the left hand side of eq.(8), eq.(30), and eq.(33). Denoting
this operator as $\Pi$, the complete one loop partition function is given
by:
$$
\Theta(\zeta,\bar\zeta)={\rm Tr}e^{2\pi i{\rm Re}\zeta\Pi}e^{-2\pi
{\rm Im}\zeta H}=
{\rm Tr}\bigl( e^{\pi i\zeta (\Pi +H)}e^{-\pi i\bar\zeta (H-\Pi)}\bigr).
\eqno(44)
$$
Using the explicit form of the Hamilton operator, eq.(26), and the form of
$\Pi$ given by eq.(33), we see, as a novel feature, that both $H+\Pi$ and
$H-\Pi$ contain both the left and the right moving oscillators. Indeed as
$$
H=L_0+\bar L_0+\hat a,\qquad \Pi=L_0-\bar L_0-{P\over2}Q_\alpha,
$$
where the chiral $L_0$ ($\bar L_0$) operators are the zero index members
of the two \lq\lq light-cone" Virasoro algebras built from $W_n$
(respectively $\bar W_n$) and from
the chiral
parts of the $x^A$ oscillators:
$$
L_n=W_n+{1\over2}\sum\limits_m:\tilde\alpha_{n-m}^{-A}
\tilde\alpha_{m}^{-A}:\   ;
$$
one can see that here -- in contrast to the usual situation -- $\Pi$
is not just the difference between the left and right null Virasoro
generators. As a consequence both $H-\Pi$ and $H+\Pi$ contain the
operator ${P\over2}Q_\alpha$ that depends on both type of light cone
oscillators. Therefore,
if we denote the trace over one set of zero mode, one set of
$n+P/2$ modded and one set of $n-P/2$ modded
oscillators in the combination given by eq.(44) as
$$
\chi(q,\bar q)={1\over (1-(q\bar q)^{P/4})}\prod\limits_{n=1}^\infty
(1-q^n(q\bar q)^{P/4})^{-1}(1-q^n(q\bar q)^{-P/4})^{-1},
\eqno(45)
$$
($q=e^{2\pi i\zeta}$), then, for fixed $P$, the one loop partition function
takes the form:
$$
\Theta(\zeta,\bar\zeta)={\eta (q)^{-d}\bar\eta (q)^{-d}\over
({\rm Im}\zeta)^{d/2}}
(q\bar q)^{{1\over2}(-bP^2-{2\over12}+1+{P\over2}(1-{P\over2}))}
\chi(q,\bar q)\bar\chi(q,\bar q).
\eqno(46)
$$
(In eq.(46) $\eta (q)$ denotes the Dedekind function). We emphasize that in
deriving eq.(46) we encountered no divergent sums and correspondingly no
regularization problem unlike in the analogue computation in the covariant
quantization [5]. We obtain this nice feature because the non vanishing
$P$ acts as a regulator in eq.(45), however in the $P\rightarrow 0$ limit
$\chi (q,\bar q)$ becomes singular: $\chi (q,\bar q)\mapsto 1/(P\pi
{\rm Im}\zeta)\prod (1-q^n)^{-2}$.

It is easy to see that
$\Theta(\zeta,\bar\zeta)$ is invariant under the $\zeta\rightarrow\zeta+1$
transformation. However the other generating transformation of the modular
group, $\zeta\rightarrow -{1\over\zeta}$, does not leave $\Theta$ invariant.
Intuitively it is easy to understand
the reason behind this : this transformation exchanges
the $\tau$ and $\sigma$ directions and the fields described by eq.(21,22)
exhibit different boundary conditions in these directions: they are periodic
in $\sigma$ but in $\tau$, because of the $e^{\pm i\Pt}$ factors, they are
only quasiperiodic.

Since modular invariance is a crucial requirement for a consistent string
theory, we also derived the contribution of the $x^i$ coordinates to
the one loop partition function using the Euclidean functional integral
approach, where the boundary conditions appear in an alternative form.

After continuing the time coordinate $\tau\to -iT$ in the light cone form of
the action, eq.(9), and introducing the
$$
z=\sigma+iT,\quad
\bar z=\sigma-iT,\quad \zeta=\tau_1+i\tau_2
$$
parametrization of the world sheet torus, we used
the complete system of functions:
$$
\psi_{(n,m)}(z,\bar z)=\exp\Bigl({\pi\over\tau_2}\bigl(n(z-\bar z)
+m(\zeta\bar z-\bar\zeta z)\bigr)\Bigr),
$$
periodic in both directions: $\psi_{(n,m)}(z+1,\bar z+1)=
\psi_{(n,m)}(z,\bar z)$; $\psi_{(n,m)}(z+\zeta,\bar z+\bar\zeta)=
\psi_{(n,m)}(z,\bar z)$ and
the standard procedure of $\zeta$ function regularization (see e.g.
[10]),
to evaluate the determinant ${\rm det}({\cal D}^{(+)}
{\cal D}^{(-)})$, where
$$
{\cal D}^{(\pm)}=-4\partial_z\partial_{\bar z}+P^2\pi^2\pm 2\pi iP
(\partial_z+\partial_{\bar z}).
$$
This way we got
$$
\bigl({\rm det}({\cal D}^{(+)}{\cal D}^{(-)})\bigr)^{-1/2}=
(q\bar q)^{{P\over4}-{1\over12}}\chi(q,\bar q)\bar\chi(q,\bar q)
\eqno(47)$$
for the $x^i$-s
partition function.
Since eq.(47) was obtained by using manifestly periodic boundary conditions,
and it reproduces (apart from a not very interesting factor) the contribution
of the $x^i$ coordinates to eq.(45-46), we conclude, that the modular non
invariance of $\Theta (\zeta,\bar\zeta)$ in eq.(46) is not the result of
choosing improper boundary conditions.

Possibly, to obtain a modular invariant partition function,
- especially the one suggested in ref.[5], - one
should introduce some sort of \lq\lq winding (and/or twisted) sectors"
conceivably together with
some appropriate projections. For these new sectors we need new ground
states in the operator formalism, or appropriate \lq\lq instanton"
solutions in the functional integral formalism. A detailed investigation
of this problem is beyond the scope of the present paper and we merely note
that a candidate for the new
ground state is provided by
$$
\bigl(x^1\pm ix^2\bigr)(\sigma)=
\exp \bigl(\mp {iP\over2}(\sigma -\sigma_0)\bigr),
$$
giving a static solution of eq.(6). It is straightforward to show that
this solution, (just like the $x^i\equiv 0$ ground state used so far),
 - when augumented by $x^A\equiv 0$ - saturates the lower
bound, ($-kbP^2/2$), of the Hamiltonian, eq.(11). As it stands, this
solution is not periodic on $(0,2\pi)$, nevertheless it's form guarantees,
that no surface term arises in the case of {\sl periodic} variations.
It is also worth pointing out that this solution is not invariant
 under separate $\sigma $ translations and $x^i$ rotations, however
 an appropriate
 combination of these global symmetry
transformations does leave it invariant.

\centerline{\underbar{8. Conclusions}}
\bigskip

In this paper we reviewed the main points of the quantization of the
Nappi Witten string in the light cone gauge. It was shown, that imposing
this gauge in an appropriately transformed form of the Nappi Witten
background linearizes the equations of motion for the relevant
degrees of freedom. Analyzing and quantizing these equations we
exhibited the presence of two sets of twisted oscillators in both
light-cone directions,
deduced the energy and mass spectra and uncovered the global
symmetries. Among the novel features it was
found that the string's
CM coordinate executes an oscillatory motion in the two transverse
directions that correspond to the spatial part of the Nappi Witten
metric. In addition it was also pointed out that the operators
corresponding to these CM oscillations also act as the step operators
of the global $E_2^c\times E_2^c$ symmetry algebra.
It was demonstrated and explained why the critical dimension
 cannot be obtained from the usual argument in the operator formalism
used in the present paper.
We also showed, that the limitations on the range of the $t$ parameter,
found in the covariant framework
from the absence of negative norm states, follow also in the light cone
formalism if we insist on the existence of a single ground state.

\centerline{\bf Acknowledgements}
\bigskip

Three of us (P.F, Z.H and L.P) would like to thank Tours University
for the hospitality and the Hungarian National Science and Research
Foundation (Grant No. 2177) for a partial financial support.

\centerline{\bf References}
\bigskip

\item{[1]} D. Amati and C. Klimcik, {\sl Phys. Lett.} {\bf B219},
443 (1989); H. de Vega and N. Sanchez, {\sl Nucl. Phys.} {\bf B317},
706 (1989); A.A. Tseytlin, {\sl Phys. Lett.} {\bf B288}, 279 (1992);
{\sl Nucl. Phys.} {\bf B390}, 153 (1993); {\sl Phys. Rev.} {\bf D47},
3421 (1993); C. Duval, Z. Horv\'ath and P.A. Horv\'athy, {\sl Phys.
Lett.} {\bf B313}, 10 (1993); E.A. Bergshoeff, R. Kallosh, T. Ortin,
{\sl Phys. Rev.} {\bf D47}, 5444 (1993); R.E. Kallosh, A.D. Linde, T.M.
Ortin, A.W. Peet, A. van Proyen, {\sl Phys. Rev.} {\bf D46}, 5278
(1992); K. Sfetsos, {\sl Phys. Lett.} {\bf B324},
335 (1994); K. Sfetsos and A. Tseytlin, {\sl Nucl. Phys.} {\bf B427},
245 (1994); C. Klimcik and A. Tseytlin, {\sl Phys. Lett.} {\bf B323},
305 (1994).

\item{[2]} G.T. Horowitz and A.R. Steif, {\sl Phys. Rev.} {\bf D42},
1950 (1990); A. Steif, {\sl Phys. Rev.} {\bf D42}, 2150 (1990).

\item{[3]} C. Nappi and E. Witten, {\sl Phys. Rev. Lett.} {\bf 71},
3751 (1993).

\item{[4]} E. Kiritsis and C. Kounnas, {\sl Phys. Lett.} {\bf B320},
264 (1994).

\item{[5]} E. Kiritsis, C. Kounnas and D. L\"ust, {\sl Phys. Lett.}
{\bf B331}, 321 (1994).

\item{[6]} P. Goddard and C.B. Thorn, {\sl Phys. Lett.} {\bf B40},
235 (1972).

\item{[7]} R. Hwang {\sl Nucl. Phys.} {\bf B354}, 100 (1991); J. Balog, L.
O'Raifeartaigh, P. Forg\'acs and A. Wipf, {\sl Nucl. Phys.}
{\bf B325}, 225 (1989).

\item{[8]} C.B. Thorn, {\sl Nucl. Phys.} {\bf B286}, 61 (1987).

\item{[9]} P.Forg\'acs, Z. Horv\'ath and L. Palla, unpublished

\item{[10]} P. Ginsparg, {\bf In} \lq\lq Fields, strings and Critical
Phenomena" ed. by E. Br\'ezin, and J. Zinn-Justin, Elsevier Science
Publishers, B. V. (1989).

\item{[11]} R.E. Rudd, {\sl Nucl. Phys.} {\bf B427}, 81 (1994);
\bye